%% file: main.tex
\documentclass[11pt]{article}
\usepackage[final]{acl}

\usepackage{times}
\usepackage{latexsym}
\usepackage[T1]{fontenc}
\usepackage[utf8]{inputenc}
\usepackage{amsmath,amssymb,amsfonts}
\usepackage{algorithmic}
\usepackage{graphicx}
\usepackage{textcomp}
\usepackage{xcolor}
\usepackage{subcaption}
\usepackage{adjustbox}
\usepackage[most]{tcolorbox}
\usepackage{multirow}
\usepackage{wrapfig}
\usepackage{enumitem}
\usepackage{tabularx}

\definecolor{lightgray}{RGB}{240, 240, 240}
\newtcolorbox{textbox}{
    colback=lightgray,%
    colframe=white
}
\newtcolorbox{quotebox}{
    colback=white,%
    colframe=white
}

\def\BibTeX{{\rm B\kern-.05em{\sc i\kern-.025em b}\kern-.08em
    T\kern-.1667em\lower.7ex\hbox{E}\kern-.125emX}}

\usepackage{xspace}

\newcommand{\ds}{Deepseek Coder\xspace}
\newcommand{\qw}{Qwen2.5-Coder\xspace} 
\newcommand{\stc}{StarCoder2\xspace}

\usepackage{booktabs}
\begin{document}

\title{
Understanding Secret Leakage Risks in Code LLMs: A Tokenization Perspective}
\author{
  Meifang Chen$^{\dagger}$,
  Zhe Yang$^{\ddagger}$,
  Huang Nianchen$^{\S}$, \\
  \textbf{Yizhan Huang}$^{\dagger}$\thanks{Corresponding author.},
  \textbf{Yichen Li}$^{\dagger}$,
  \textbf{Zihan Li}$^{\P}$, and
  \textbf{Michael R. Lyu}$^{\dagger}$ \\
  $^{\dagger}$The Chinese University of Hong Kong   \quad \quad $^{\ddagger}$Nanyang Technological University\\
  $^{\S}$University of Southern California   \quad \quad $^{\P}$Fudan University\\
  \texttt{yzhuang22@cse.cuhk.edu.hk}
}

\maketitle

\begin{abstract}
Code secrets are sensitive assets for software developers, and their leakage poses significant cybersecurity risks. While the rapid development of AI code assistants powered by Code Large Language Models (CLLMs), CLLMs are shown to inadvertently leak such secrets due to a notorious memorization phenomenon. This study first reveals that Byte-Pair Encoding (BPE) tokenization leads to unexpected behavior of secret memorization, which we term as \textit{gibberish bias}. Specifically, we identified that some secrets are among the easiest for CLLMs to memorize. These secrets yield high character-level entropy, but low token-level entropy. 
Then, this paper supports the biased claim with numerical data. We identified that the roots of the bias are the token distribution shift between the CLLM training data and the secret data. We further discuss how gibberish bias manifests under the ``larger vocabulary'' trend. To conclude the paper, we discuss potential mitigation strategies and the broader implications on current tokenizer design. 
\end{abstract}

\input{sections/intro}
\input{sections/bg}

\input{sections/pilot_study.tex}

\input{sections/main_exp}
\input{sections/relate}
\input{sections/discuss}

\input{sections/ack.tex}

\bibliography{references}
\input{sections/appendix}

\end{document}

%% file: sections/intro.tex
\section{Introduction}
\begin{figure*}[t]
    \centering
    \includegraphics[width=1\linewidth]{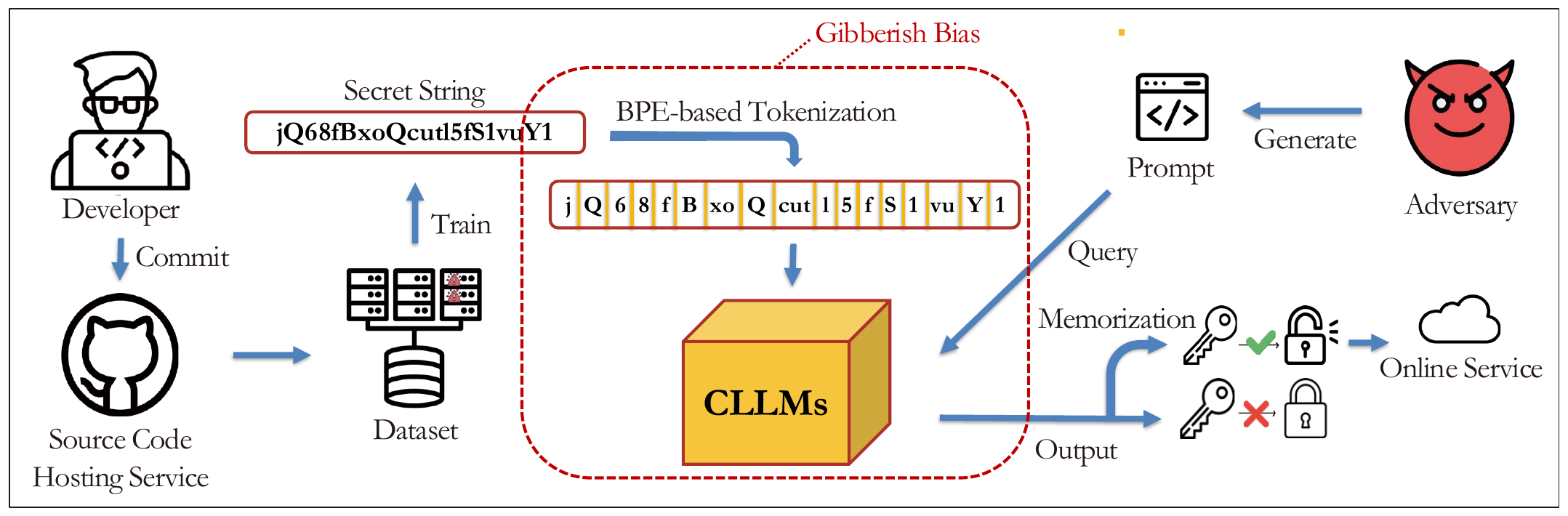}
    \caption{The risk map of secret leakage through CLLMs. The red box highlights the focus of this paper.}
    \label{fig:risk}
\end{figure*}
Code Large Language Models (CLLMs) are reshaping how software is built, maintained, and evolved, introducing AI-driven automation across every stage of the software development lifecycle.  
Recent academic advances \cite{hui2024qwen,huggingface2024starcoder2,guo2024deepseek}, demonstrate significant improvements in code generation, understanding, and reasoning across multiple programming languages and tasks. 
These models have been evaluated on challenging benchmarks showing state‑of‑the‑art performance in code synthesis, completion, and bug repairs. On the commercial side, tools like Claude Code~\cite{claude_code_web}, Codex~\cite{codex_web}, and Cursor~\cite{Cursor} have operationalized CLLMs by integrating the technology directly into real-world development environments to provide real‑time coding assistance, documentation generation, and automated testing. Such tools have achieved rapid market penetration; for instance, Claude Code has seen swift developer adoption, quickly accumulating over 111,000 cumulative npm downloads by early 2026~\cite{GraduallyAI_ClaudeCode_2026}.

However, the broad adoption of CLLMs is increasingly raising concerns in the community. Code LLMs are shown to have strong capability of \textit{memorization} -- with proper prompts, they emit training data verbatim. For instance, CLLMs may leak documentations, statements, logs, and configuration files~\cite{unveil,wu2025empirical,pearce2022asleep}. CLLMs may even memorize and leak code \textit{secrets}.  
These secrets include API keys to online services, private keys, passwords, and URLs, posing severe security and privacy risks. The secrets appear in training corpus, since careless programmers push their secrets to code hosting services like GitHub.
Therefore, with the exploding use of CLLMs, the topic is becoming important in the AI safety community. We illustrate the risk in Fig.~\ref{fig:risk}.

While most memorization works focus on prompting, training paradigms, and datasets of CLLMs, this paper explores tokenization, an under-explored yet pivotal component of CLLMs. Our research is motivated by the recently discovered relationship  between entropy and memorization score. The work reveals that for a token sequence from the LLM training corpus, (an estimator) of entropy is linearly related to memorization score~\cite{huang2025entropy}. By applying the previous discovery to the memorization of secrets, this study
reveals that although some gibberish-like secrets (i.e., highly-randomized strings) are \textit{high}-entropy at the character-level, after tokenization, some secrets are encoded to \textit{low}-entropy token sequences, significantly reducing chances of memorization. We coin the phenomenon as \textbf{gibberish bias}.  Our research results indicate that the risk of secret leakage is exacerbated with gibberish bias.

This study conducts a deeper exploration of gibberish bias. We found that gibberish bias should be attributed to Byte-pair Encoding (BPE), the most popular tokenization strategy in (Code) LLMs. We confirm that the root of gibberish bias --- BPE is sensitive to distribution shift between \textit{train} data and inference (\textit{test}) data well. 
We then discuss the bias under the current trend of ``larger tokenizers''. The final part of the paper discusses the mitigation strategy and its broader implications for tokenizer design. The contribution of this paper is as follows:

\begin{enumerate}[leftmargin=*]
    \item This paper identifies a new risk of secret leakage through CLLMs: the tokenizer might induce gibberish bias and further exacerbate the secret leakage risk.
    \item This paper explores the roots of such bias: BPE is sensitive to the distribution shift between train and test data.
    \item This paper predicts that the secret leakage risk will manifest more under the current ``larger tokenizer'' trend.
    \item The paper discusses potential mitigation strategies and broader implications for the community.
\end{enumerate}

%% file: sections/bg.tex
\section{Background}
\subsection{Tokenization of Code LLMs}
\label{bg_tokenization}
\begin{figure}[htbp]
    \centering
    \includegraphics[width=\linewidth]{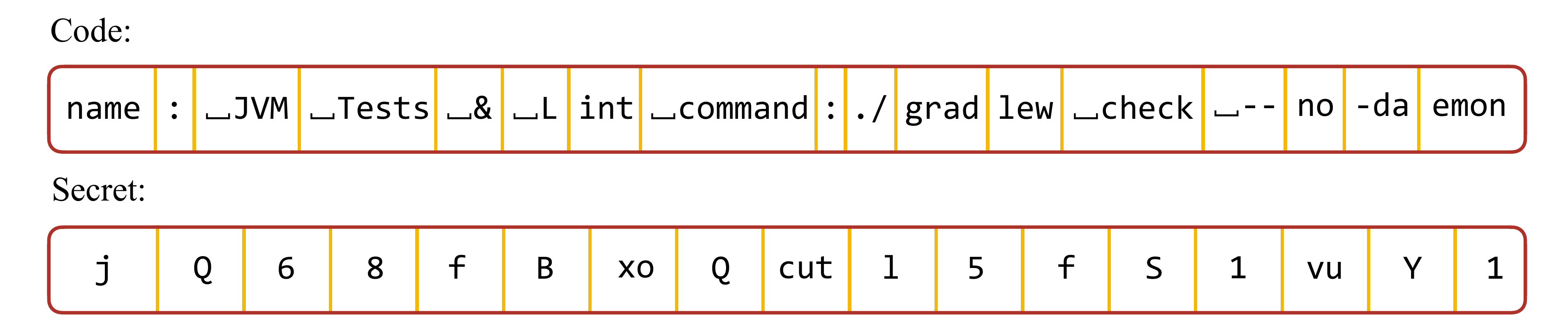}
    \caption{Use BPE to tokenize one line of normal code and secrets. Each token is a subword.}
    \label{fig:tokenization_case}
\end{figure}
In NLP, early explorations use word-level tokenization (e.g., word2vec~\cite{mikolov2013efficientestimationwordrepresentations}), assigning each distinct word a unique index and embedding. This approach imposed a fixed vocabulary and sometimes resulted in the failure to handle out-of-vocabulary (OOV) words~\cite{mikolov2013efficientestimationwordrepresentations}. Later, researchers developed subword tokenization strategies that segment words into smaller units. 

The process typically includes three sequential stages: 
\textit{(1) Pre-tokenization:} a preprocessing step that imposes rules on the raw text, such as whitespace splitting, normalization, or restrictions on allowable character sequences.
\textit{(2) Vocabulary Construction:} given a corpus and a target vocabulary size, an algorithm selects a set of subword units that constitute the vocabulary,
Each subword is regarded as a \textit{token}.
\textit{(3) Segmentation:} using the constructed vocabulary, this step operationalizes the mapping from raw text to subword sequences. 

We note that before LLM training, model developers run stages (1) and (2) to establish the vocabulary. The vocabulary should be strictly adhered to all stages of LLM training (i.e., pre-training, post-training). During LLM training and inference, only stages (1) and (3) are applied. In decoder-only LLMs, a tokenizer assigns an input string to a sequence of numerical IDs, which are further transformed into  embeddings via an embedding matrix.

The core of tokenization is the vocabulary construction algorithm (stage (2)), and the Byte-Pair Encoding (BPE)~\cite{sennrich-etal-2016-neural, bpe_origin} is the de facto standard for LLMs since GPT-2~\cite{gpt2}. It is a greedy, data-driven merge algorithm that iteratively combines the most frequent adjacent character or byte pairs from the training corpus to build a subword vocabulary. Starting from individual symbols, BPE yields multi-character tokens for frequent patterns (e.g., “ing”, “tion”) while splitting rare words into smaller units. The total merging step is a hyper-parameter, and depends on the empirical decision of model trainers.

Code LLMs, including \ds~\cite{guo2024deepseek}, \stc~\cite{huggingface2024starcoder2}, \qw~\cite{hui2024qwen}, generally employ BPE-based tokenization, following the tokenizers on LLMs. For CLLM tokenizers, the vocabularies are slightly adapted towards code-related tasks. For example, many Code LLMs adapt vocabularies on codebases to improve generation fidelity and ensure compilable output~\cite{wang-etal-2021-codet5}. 
We further showcase how tokenization works on an example string in Fig.~\ref{fig:tokenization_case}.
\subsection{Code Secrets and their Entropy}
\label{bg_secret}
Software developers need \textit{secrets}  to authenticate these third-party services as
part of system integration. The secrets include API keys, access tokens, and private keys. While secrets are sensitive assets during software development, careless developers may hard-code secrets in their code and push it to online code hosting services like GitHub. Recent studies have shown that a vast amount of secrets are exposed in public software repositories~\cite{secretbench,passfinder,meli2019bad,10.1145/3576915.3616591,ndss2024skeleton}. These secrets are collected as part of the training corpus of CLLMs; hence, they might be accidentally leaked by CLLMs. 
\begin{table}[h]
\small
\centering

\caption{Common secrets and their corresponding regex patterns.}
\begin{tabular}{l|l}
\hline
\textbf{Secret type} & \textbf{Regex} \\
\hline
AWS Access Key ID & \verb|AKIA[0-9A-Z]{16}| \\
Google API Key & \verb|AIza[0-9A-Za-z-_]{35}| \\
Tencent Cloud Secret ID & \verb|AKID[0-9a-zA-Z]{32}| \\

\hline
\end{tabular}
\label{tab:secret_regex}
\end{table}

An important property of code secrets is that secrets typically exhibit high (char-level) entropy~\cite{shannon2001mathematical}.  For a human being, the strings look like gibberish.  High entropy indicates high uniqueness. Consequently, for online service providers, entropy is a pivotal design consideration for the security of their secrets. Secrets typically follow a specific format, hence can be characterized by regular expressions. Table ~\ref{tab:secret_regex} presents examples of secrets of popular online services.

Mathematically, denote a secret as a sequence  $s=(s^1,s^2,...,s^{|s|})$, where each atomic element could be either a char or a token. Denote the set of all possible outcomes as $\mathcal{V}$. Denote the expected frequency of $x$ as $p(x)$,  the entropy $H$ of secret $s$ is 
\begin{equation}
  H(s)\triangleq  - \sum_{x \in  \mathcal{V} } p(x) \log p(x).
  \label{eq:entropy}
\end{equation}

Taking GitHub personal access tokens \footnote{``Tokens'' in ``personal access tokens'' refers to the strings for authentication purposes. Readers shall not confuse with ``tokens'' used in (C)LLM tokenization.}
~\cite{harvey_2021} as an example, such tokens follow the format specified by the regular expression \verb|ghp_[a-zA-Z0-9]{36}|. The char-level entropy for this token is 5.915 bits, which is close to the maximal entropy of 5.977 bits. We defer the detailed calculation to appendix~\ref{appdx}.

\section{Motivating Study: BPE Is \textit{Counterintuitive} on Secret Tokenization}
Secrets are designed at character-level -- the regular expression defines the secret format, and the randomized part is random \textit{characters}. However, CLLMs process these strings on the token-level.  This section then presents a motivating study using visualizations, and reveals that BPE indeed results in \textit{counterintuitive} behaviors.
\subsection{Case Study}
\label{case_study}

This case study discusses the tokenization example in Figure \ref{fig:tokenization_case}. The figure illustrates how {\ds}~\cite{guo2024deepseek} tokenizes a part of normal code and a secret (substring). We observe a significant difference in \textbf{token granularity}: Although two strings present the same token count, the source code is significantly longer measured in character-level length. 

For typical source code, the tokenizer chunks text in a way that is much expected: a token typically consists of multiple characters. Some words are split into shorter sub-words. For example, the word ``-daemon'' is represented by two tokens, ``-da'' and ``emon''. In contrast, the situation is much different for secrets.  Most of the tokens are one character. Other tokens are characters of length 2 and 3. Interestingly, the tokenizer identifies an English word ``cut'' in a gibberish-like string as a token.

Even with one example, we could observe how BPE-based tokenization might induce gibberish bias. Tokens of secrets are distributed in a highly non-uniform way. They include chars of length 1, 2, 3, maybe $n>10$. Tokens of shorter char-lengths may appear with higher frequency, while longer ones may be less frequent. In information theory, uniform distribution exhibits the maximal entropy possible, and hence non-uniform random variables exhibit less entropy. With less entropy, secrets are increasingly likely to be memorized. We leave the detailed calculation to Section~\ref{rq}.

\subsection{Tokenizing 2-char secret sub-strings}

\begin{figure}[t]
    \centering
    \begin{subfigure}[t]{0.49\linewidth}
        \includegraphics[width=\linewidth]{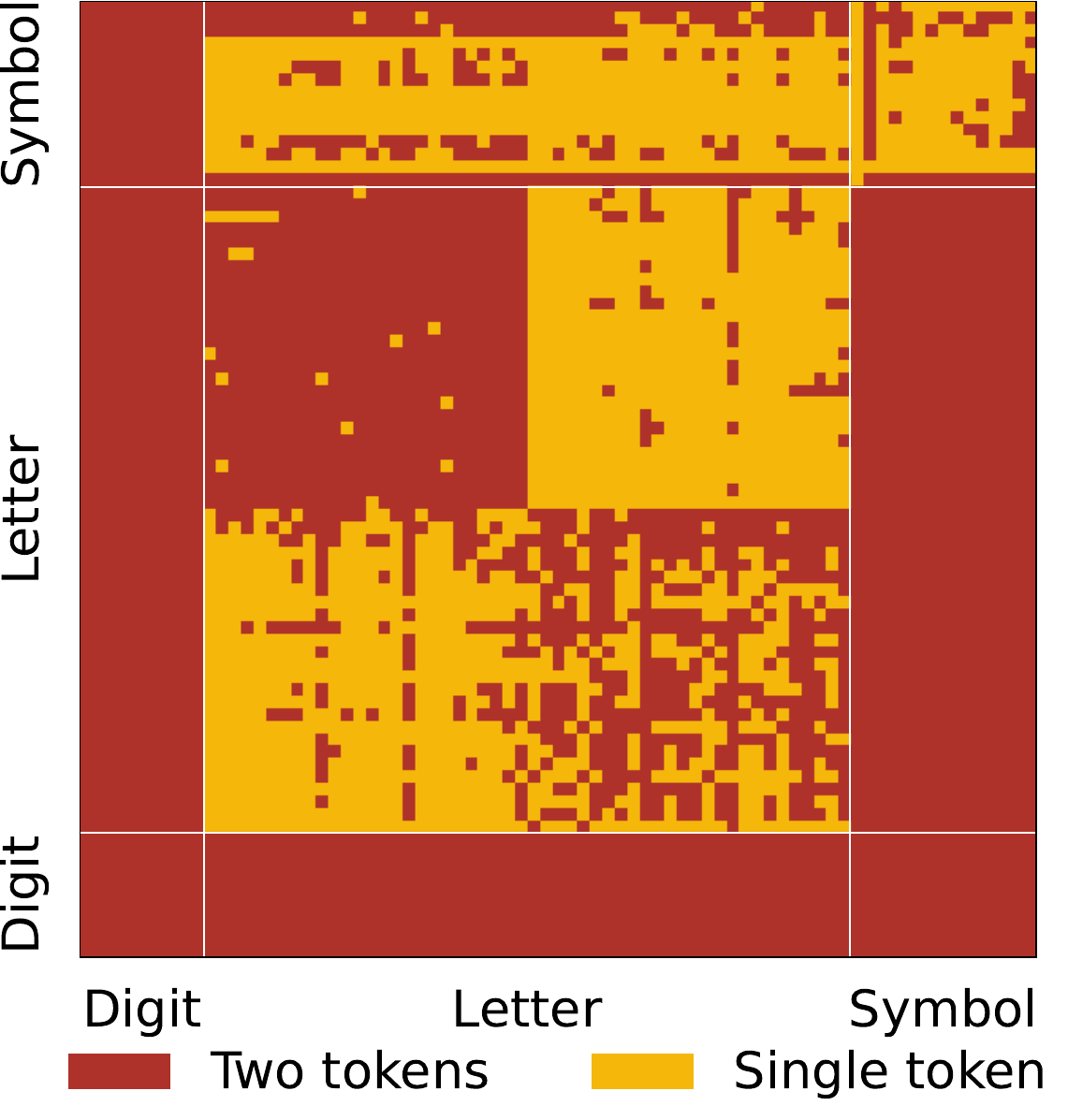}
    \end{subfigure}%
    \hfill
    \begin{subfigure}[t]{0.49\linewidth}
        \includegraphics[width=\linewidth]{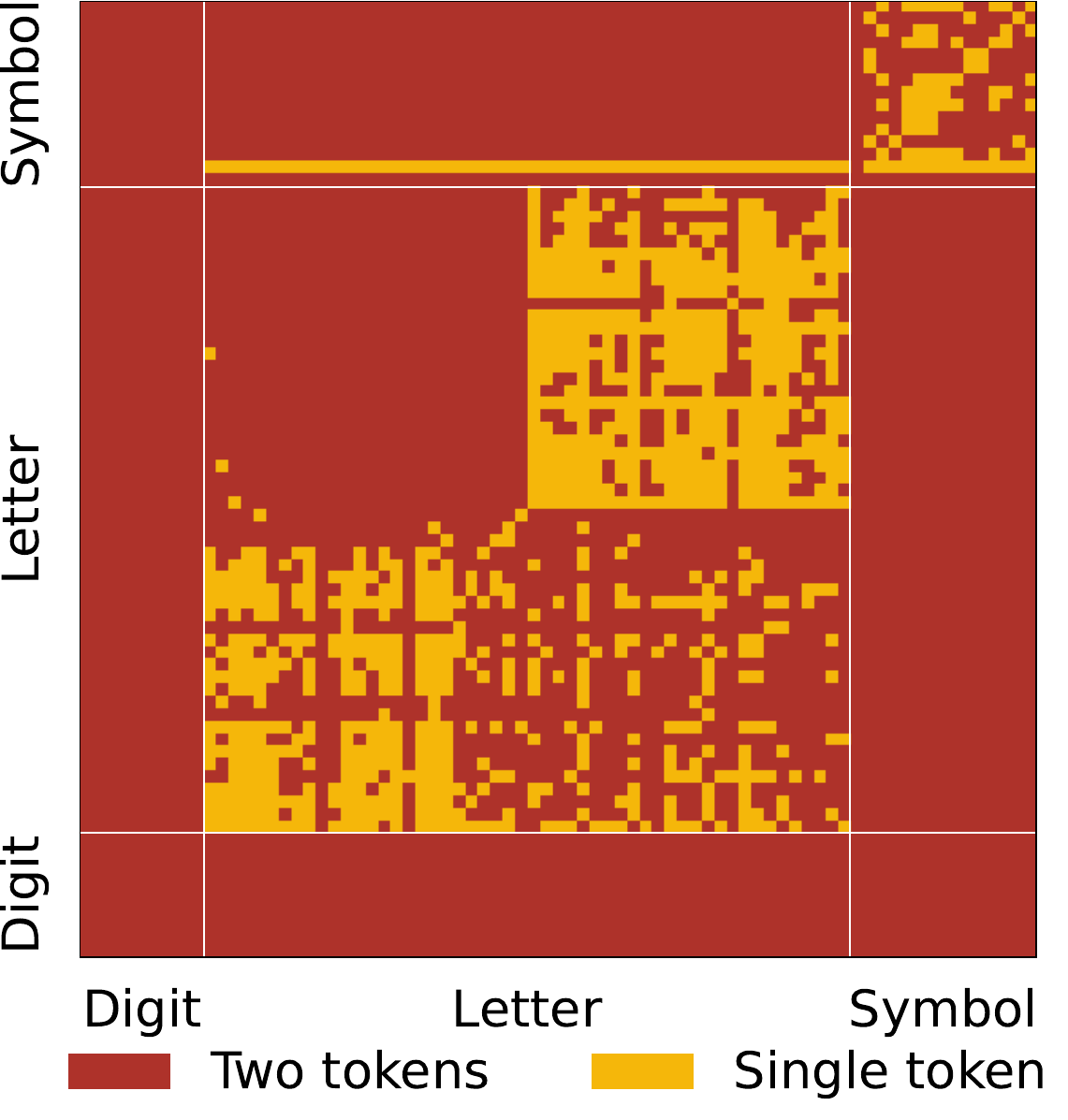}
    \end{subfigure}
    \caption{Visualizing how {\qw} (left) and {\ds} (right) tokenize a 2-char sub-string of secrets. 
    Denote the substring as $a_1a_2$, $a_1$ corresponds to the horizontal coordinates of a pixel, 
    and $a_2$ corresponds to the vertical coordinates of a pixel.}
    \label{fig:2-char}
\end{figure}

 To demonstrate how tokenizers perform poorly on secrets, this section shows the tokenization of a 2-char randomly-generated string. The characters are selected from a vocabulary of size 76, i.e., letters (a-z, A-Z), digits (0-9), and other symbols. These chars are commonly used by secrets. We enumerate all the combinations of a 2-char string given the vocabulary. We examine each string whether it is encoded to \textit{one} token (in yellow) or \textit{two} tokens (in brown) and present the result in Figure \ref{fig:2-char}. There is clearly a non-random distribution of different encoding strategies. In general, we observe certain character pairs that the tokenizer tends not to split: (lowercase letter, lowercase letter), (uppercase letter, uppercase letter), (symbol, symbol), (letter, symbol). However, as one easily finds out, many non-trivial corner cases exist. The overall decision boundary is thus governed by a complex interplay of statistical heuristics. Similar patterns are observed on the {\ds} tokenizer. We further observe consistent patterns on unigram-based tokenizers (XLNet, T5); see Appendix~\ref{appdx:unigram_heatmap}.

%% file: sections/pilot_study.tex
\section{Gibberish Bias} 
\label{main}
Section~\ref{case_study} demonstrated that on secrets, BPE-based tokenization may not function as ideally as on normal data. In other words, it is potentially biased -- we term it \textbf{gibberish bias}. This section aims to formally describe gibberish bias using mathematical notations. Inspired by several research works on secret memorization, we characterize gibberish bias by memorization.  Secret leakage has strong implications for the community, since upon leakage, these credentials may grant an adversary access to online services, bringing severe cybersecurity concerns.

\subsection{Preliminaries}
A recent LLM memorization work~\cite{huang2025entropy} has shown that entropy, the essential property that comes with the design of code secrets, is closely related to memorization. The work discovers the so-called \textit{Entropy-Memorization Law}, which could be formally described as follows.

Denote a fixed pre-trained LLM $\theta$, prompt (a token sequence) $p$, the golden answer (a token sequence) $s$, and a memorization score measuring the difference between LLM continuation and the golden answer $d(\theta(p),s)$. 

The work studies two metrics, \textbf{entropy} and \textbf{normalized entropy} of the answer sequence, $M(s)$, and $\overline{M}(s)$. Note that the entropy notation is slightly twisted from $H(s)$ in Equation~\ref{eq:entropy}. The subtle difference lies in the outcome space $\mathcal{V}$. Since LLMs process strings at the token-level, $\mathcal{V}$ is defined over tokens. Moreover, in the Entropy-Memorization Law, the authors adopted a level-set-based calculation for $\mathcal{V}$, which we skip due to space limitations. 

With the established entropy notation, normalized entropy quantifies how closely the entropy of a sequence approaches the maximum possible entropy. Normalized entropy eliminates the effects of sample space size. It is defined as
\begin{equation}
    \overline{M}(s) \triangleq  \frac{M(s)}{M_{\max}(s)} = \frac{M(s)}{\log |\mathcal{V}|}.
    \label{eq:normalize}
\end{equation}

Under specific conditions, a loose statement of the Entropy-Memorization Law is:
\begin{textbox}
1. $M(s)$ is a proxy of $d(\theta(p),s)$. The relation is positively linear.

2. $\overline{M}(s)$ is a proxy of $d(\theta(p),s)$. The relation is negatively linear.
\end{textbox}
 In other words, higher entropy of a token sequence indicates a lower chance of memorization. Higher normalized entropy of a token sequence indicates a higher chance of memorization.

 In this work, we are interested in studying the following metrics of the secret strings:  character-level entropy $M(s)$, token-level entropy $H(s)$, and the normalized entropy, and their normalized version $\overline{M}(s)$, $\overline{H}(s)$.

\subsection{Tokenizing secrets}
\paragraph{Experimental Setup.}
We adopt the same experimental setup as \cite{huang2025entropy}. We use OLMo-1B~\cite{olmo1b}, a fully-open LLM with its training corpus Dolma~\cite{dolma}.  In the experiments, we sampled 240k sequences from Dolma. The dataset is licensed by ODC-BY, granting free  access for research purposes. We reproduce the results with the same algorithm adopted in \cite{huang2025entropy}. We study the zero-distance set (where memorization score is 0, or ``perfect memorization'') and try to identify secret strings.

\paragraph{Experimental Results.}

Among all 847 instances within the zero-distance set, we identified 102 gibberish-like secrets through manual labeling. Some  examples are shown below.  

\definecolor{codegreen}{rgb}{0,0.6,0}
\definecolor{codegray}{rgb}{0.5,0.5,0.5}
\definecolor{codepurple}{rgb}{0.58,0,0.82}
\definecolor{backcolour}{rgb}{0.95,0.95,0.95}
\lstset{
    backgroundcolor=\color{backcolour},   
    numberstyle=\tiny\color{codegray},
    stringstyle=\color{codepurple},
    basicstyle=\ttfamily\footnotesize,
    breakatwhitespace=false,
    breaklines=false,
    captionpos=b,                    
    keepspaces=true,                 
    numbers=left,                    
    numbersep=5pt,                  
    showspaces=false,                
    showstringspaces=false,
    showtabs=false,                  
    tabsize=2
}
\begin{lstlisting}[
caption= Three examples of gibberish generated by OLMo-1B. Part of these secrets are masked due to privacy concerns.,
numberstyle=\small\ttfamily\color{black},
xleftmargin=2em]
ESVXlO3url***Dw23KjZ
PszYzqs83S***N2N0Rj
5laXo6c1Ib***stxU9
\end{lstlisting}

We conduct analysis over four sets: gibberish, non-gibberish, zero-distance set, and non-gibberish in the zero distance set~\footnote{To clarify, non-secrets refers to the complement set of \textit{labeled} secrets. Therefore, ``non-secrets''  may include unlabeled secrets. Due to the large size of the sampled corpus,  labeling on such an extensive scale is not feasible at this stage of work.}. We adopt the entropy estimator and normalized entropy introduced in this section over these four sets. The results of our analysis are summarized in the following table.

\begin{table*}[h]
\small
\centering
\caption{Statistics of secret memorization at token-level (\textit{T}) and char-level (\textit{C}).}
\resizebox{\textwidth}{!}{%
\begin{tabular}{lcc|cc|cc}
\toprule
\multirow{2}{*}{} 
& \multicolumn{2}{c|}{\textbf{Unique Elements}} 
& \multicolumn{2}{c|}{\textbf{Entropy}} 
& \multicolumn{2}{c}{\textbf{Normalized Entropy}} \\
& \textit{T} & \textit{C} 
& \textit{T} & \textit{C} 
& \textit{T} & \textit{C} \\
\midrule
\textbf{Zero-distance Set} & 3,661 & 105 & 7.834 & 5.110 & 0.662 & 0.761 \\
\textbf{Secrets}         & 1,047 & 76  & 8.084 & 6.086 & 0.806 & 0.974 \\
\textbf{Non-Secrets}     & 47,945 & 9,006 & 11.175 & 4.744 & 0.719 & 0.361 \\
\textbf{Non-Secrets in Zero-distance Set} & 2,897 & 105 & 7.329 & 4.966 & 0.637 & 0.740 \\
\bottomrule
\end{tabular}%
}
\end{table*}
The experimental results reveal the key findings: \textbf{high \textit{character}-level entropy does not necessarily imply high \textit{token}-level entropy.} In fact, at the token-level, these secrets have significantly lower entropy (8.084) than non-secrets (11.175); while at the char-level, these secrets have significantly higher entropy (6.086) than non-secrets (4.744). Besides, if we calculate the difference delta between secrets and non-secrets, there is a significant gap in normalized entropy between token-level ($\Delta=0.806-0.719=0.087)$ and character-level ($\Delta=0.974-0.361=0.613$). 

It is observed that at the token-level, the 102 identified gibberish uses more than 1k different tokens. On the contrary, at char-level, they are composed of 76 unique chars (i.e., A-Z, a-z, 0-9, and 14 other chars). That showcases the outcome space size discrepancy between char-level and token-level, and explains our finding.

The findings may deviate from our human intuition. A wrong logic chain of a human is: by the Entropy-Memorization Law, secret strings are highly randomized; hence, they have high entropy and are hard to memorize. The problem lies in the ``high entropy property'' naturally assumed by our human beings --  human beings perceive secrets at the character level. In contrast, for LLMs, the entropy should be calculated over the token level. After tokenization, some high character-level normalized entropy strings are transformed into low entropy ones.
Hence, the EM-Law suggests they should be easier to memorize than an average non-gibberish text.

 We conclude this section with the full description of gibberish bias:

\begin{tcolorbox}[title = {Gibberish bias},
  fonttitle = \bfseries]
  BPE-based tokenization transforms some of the \textbf{high \textit{character}-level} entropy sequences into \textbf{low \textit{token}-level} entropy sequences.
\end{tcolorbox}

%% file: sections/main_exp.tex
\section{How Does Tokenization Induce Gibberish Bias?}
\label{rq}
The above results reveal an essential defect induced by tokenization. The findings raise an intriguing and important question: How does this happen? To explore the question, this section then discusses the design issues of BPE. 
\paragraph{Experimental Setup}
The following experiments include tokenizers of three representative Code LLMs: \ds~\cite{guo2024deepseek}, \qw~\cite{hui2024qwen}, and \stc~\cite{huggingface2024starcoder2}. All model weights and tokenizers are free to access for research purposes. The experiments involve comparisons on two datasets: the ``secret'' dataset as introduced in Section \ref{main} and the subsample Stack V2 dataset~\cite{huggingface2024starcoder2}. Stack V2 is the training corpus of \stc.

LLM trainers typically construct a sample corpus that shares the distribution with the training data to construct the vocabulary.
Moreover,  tokenizer design is fixed during the whole LLM training (and inference) process.  However, in terms of distribution, secret strings are essentially different from the (Code) LLM training corpus. We regard that gibberish bias may stem from the distribution shift between secrets and a typical (Code) LLM training corpus. This also matches our intuition -- secret data are highly randomized, while the CLLM training corpus typically includes source code collected from online code hosting services. Then this section aims to answer the question: \textit{Compared with the token distribution of the code dataset to train CLLMs, how different is the distribution of tokens in secrets?}

\begin{figure*}[htbp]
    \centering
    \begin{minipage}{0.48\linewidth}
        \centering
        \includegraphics[width=\linewidth]{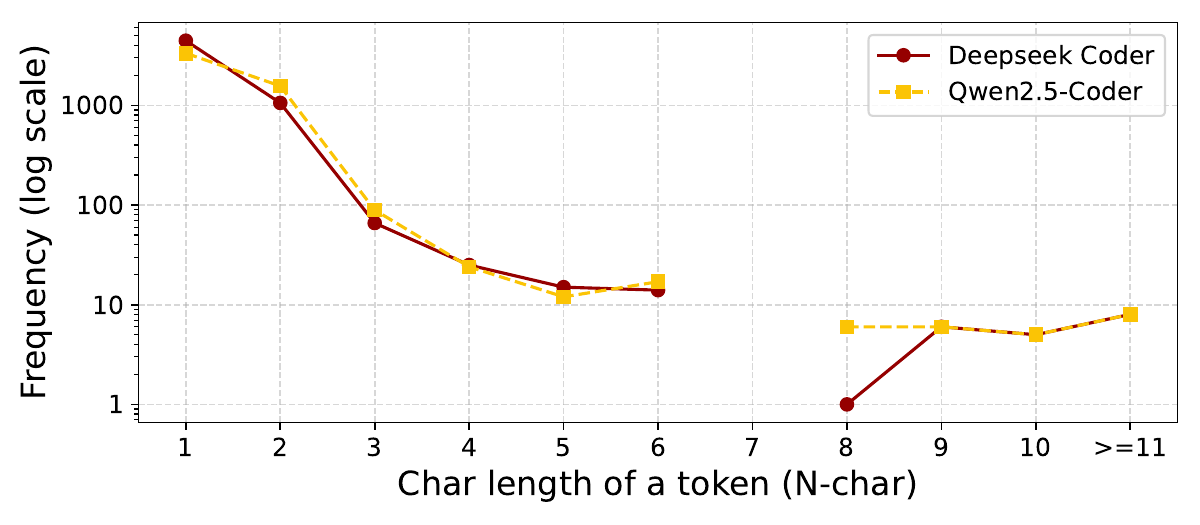}
        \caption{The distribution of n-char tokens.}
        \label{fig:n-char-count}
    \end{minipage}\hfill
    \begin{minipage}{0.48\linewidth}
        \centering
        \includegraphics[width=\linewidth]{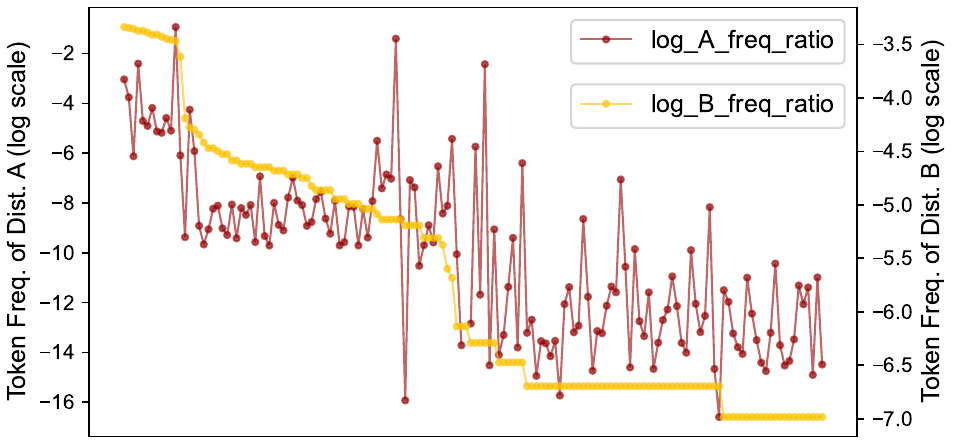}
        \caption{Token distributions on the subsampled Stack V2 dataset (Distribution A, in brown) and the secret dataset (Distribution B, in yellow). }
        \label{fig:dist_shift}
    \end{minipage}
\end{figure*}

Following the approach in the case study (Sec.~\ref{case_study}), we conduct tokenization on the secret dataset and analyze the token composition of secrets.  
To be specific, we report the frequency of tokens based on their character length ($n$) for all secret-related tokens in the dataset in Fig.~\ref{fig:n-char-count}. We observed an exponential decrease in token frequency as character length $n$ increases, as evident from the log-scale y-axis. The overall distribution is long-tailed, leading to the low entropy of the token sequence. Interestingly, we even found tokens with lengths of $n\geq11$ characters, indicating the complexity of the tokenization decision boundary. 

Ideally, the token distribution of secrets should be \textit{uniform}; but the observed distribution of BPE tokens is \textit{long-tailed}. From information theory, we know that uniform distribution achieves the maximal entropy~\cite{shannon2001mathematical}, while non-uniform distribution exhibits less entropy. The discrepancy of token distributions supports the gibberish-bias claim.

Next, we are interested in the causes of the identified long-tail distribution. How do these long char-length tokens form? We further sort the tokens on the secret dataset by frequency, and compare them with the frequency of the same token on a general CLLM dataset.

Fig.~\ref{fig:dist_shift} ranks the most frequent 150 tokens on the secret dataset, and compares the frequency of each token on the subsampled Stack V2 dataset. For both datasets, we employ the same tokenizer \stc. The tokens distribution of the secret dataset (yellow line), exhibits a much steeper and more consistent drop-off in frequency for less common tokens compared to the Stack V2 (brown line). This observed divergence in log-proportion frequencies strongly suggests a distribution shift between the two datasets, particularly for tokens beyond the most frequent few.
Setting the Stack V2 as the reference data distribution, we further report the KL divergence between the secret dataset and the Stack V2. The resulting KL divergence is 2.668, demonstrating the discrepancy.

\begin{textbox}

Using a CLLM tokenizer, the token distribution of secret data is long-tailed -- and such a long-tailed distribution explains the low entropy of secrets, hence explains the claimed gibberish bias. Further investigation confirms the significant shift in  distribution between secret data and general CLLM training data.
\end{textbox}

\section{Gibberish Bias Under a Larger Tokenizer Vocabulary}
There is a growing consensus in both academia and industry that larger models shall benefit from a larger vocabulary~\cite{tao2024scaling,huang2025overtokenized}. As the prevailing trends of building larger Code LLMs, it is believed that CLLMs deserve large vocabularies. \citet{tao2024scaling}'s work reveals that model parameters $N_{nv}$ and the corresponding optimal vocabulary size $N^{opt}_v$ approximately follow a \textit{power} law.  The authors further find that almost all state-of-the-art general-purpose LLMs and Code LLMs use vocab sizes smaller than the predicted optimal one. Then a question naturally arises: Will a larger vocabulary size induce more gibberish bias?

We build three new tokenizers of \stc (3B, 7B, and 15B) using the sub-sampled stack v2. We follow the \textit{IsoFLOPs}~\cite{tao2024scaling} approach to generate tokenizers in ``optimal'' vocabulary size. 
IsoFLOPs analysis suggests that size 39367  for the 3B model, size 62280  for the 7B model, and size 93987  for the 15B model. We then trained these tokenizers using the subsampled Stack V2 dataset, based on the suggested size. For clarity of presentations, these tokenizers are named \textit{SC-3B-o}, \textit{SC-7B-o}, and \textit{SC-15B-o}, respectively.

With three new tokenizers,  we investigate gibberish bias on the secret dataset. Following~\cite{huang2025entropy}, we report the entropy and normalized-entropy of these secrets introduced in Section~\ref{bg_secret} at the token-level.

\begin{table*}[htbp]
\centering
\caption{Entropy and normalized entropy of secrets when using different tokenizers. ``Sec.'' refers to  ``Secrets''.
}
\begin{adjustbox}{max width=\linewidth}

\begin{tabular}{@{\hskip 3pt}l@{\hskip 6pt}ccccc@{\hskip 6pt}ccccc@{\hskip 3pt}}
\toprule
 & \multicolumn{4}{c}{\textbf{Entropy}} & \multicolumn{4}{c}{\textbf{Normalized Entropy}} \\
\cmidrule(lr){2-5} \cmidrule(lr){6-9}
 & \textit{SC-3B-o} & \textit{SC-7B-o} & \textit{SC-15B-o} & \textit{Char}
 & \textit{SC-3B-o} & \textit{SC-7B-o} & \textit{SC-15B-o} & \textit{Char} \\
\midrule
\textbf{Sec.}                     & 6.931 & 7.076 & 7.145 & 6.086 & 0.777 & 0.774 & 0.773 & 0.974 \\
\textbf{Non-Sec.}                & 9.136 & 9.355 & 9.575 & 4.745 & 0.721 & 0.714 & 0.708 & 0.361 \\
\textbf{Sec./Non-Sec.}                        & 0.759 & 0.756 & 0.746 & 1.283 & 1.079 & 1.085 & 1.092 & 2.695 \\
\bottomrule
\end{tabular}
\end{adjustbox}
\label{4b}
\end{table*}

To compare metrics on secrets and non-secrets, we use a ``sec./non-sec.'' ratio. Table \ref{4b} shows the overall result. First, the results on new tokenizers converge with those of OLMo tokenizers discussed in Section~\ref{main}. Compared to non-secrets, secrets exhibit lower entropy at the token level and higher entropy at the char level; secrets achieve almost maximal normalized entropy. 
Second, regarding the sec./non-sec. ratio, it is observed that the relative difference between secrets and non-secrets is slightly expanding. For both entropy and normalized entropy metrics, the sec./non-sec. ratio keeps deviating from the baseline ratio 1. We summarize our findings in this section as follows:
\begin{textbox}

 Gibberish bias tends to be more pronounced in models employing larger tokenizer vocabularies.
\end{textbox}

%% file: sections/relate.tex
\section{Related Work}
\paragraph{Empirical quantification of secret memorization.}
Initial memorization research focused on empirically demonstrating and quantifying this risk~\cite{carlini_extracting_nodate}. HCR~\cite{huang2024your} is a method proposed to test and validate the leakage of hard-coded credentials from neural code completion tools using prompts derived from public GitHub files. Similarly, \citet{yang2024unveiling} conducted a systematic study showing that CLLMs memorize a broad range of data related to code. A study~\cite{wu2025empirical} investigates how memorization leads to general code cloning, raising concerns about copyright infringement and bug propagation. Other studies explore code clone as implementations of memorization~\cite{clone2,clone3}.
Recent research probes the underlying mechanisms at a finer granularity. For example, DESEC~\cite{nie2024decoding} addresses the problem from the decoding stage.

\paragraph{Tokenization shapes LLM performance.}
Tokenizer design has been shown to cause unexpected downstream harms, such as cross-lingual unfairness~\cite{petrov2023tokenizers} and the misalignment between subword tokenization and code grammar~\cite{tokdrift}; we identify an analogous privacy harm, showing how standard tokenizers can inadvertently reduce the token-level entropy of secrets, making them inherently more vulnerable to memorization.

%% file: sections/discuss.tex
\section{Discussions}

\subsection{Implications on Safeguarding Secrets}

\paragraph{Mitigating gibberish bias.}
We propose two mitigation strategies. The first is to enforce character-wise tokenization for secrets (details in Appendix~\ref{appdx:charwise}). The second, \textit{gibberish-token elimination}, directly targets the underlying mechanism and offers better computational efficiency. It consists of two stages: (1) \textit{identify} gibberish tokens in the tokenizer vocabulary, and (2) \textit{eliminate} them from the CLLM. The motivation is that certain tokens are used predominantly when tokenizing gibberish yet rarely appear in natural language or code; for instance, in Qwen2.5-Coder, ``aksi'' (id: 37679) and ``Ġ\textbar\textbar'' (id: 1369) do not belong to any normal programming language. We attribute such tokens to gibberish in the tokenizer's (and LLM's) training data.

\textit{Stage 1: identifying gibberish tokens.} We train two BPE tokenizers, one on a uniform subset of the CLLM corpus ($V_1$) and one on a curated secret corpus ($V_2$), and select the top-$k$ tokens that appear disproportionately often in $V_1$ but are rare or absent in $V_2$.

\textit{Stage 2: eliminating gibberish tokens.} Given the identified set, we propose two approaches. (i) Tokenizer mapping deletion removes the ``token $\rightarrow$ id'' mappings of gibberish tokens while leaving the model unchanged, preserving all parameters including the embedding matrices. (ii) Vocabulary reduction applies cross-tokenizer distillation~\cite{minixhofer_universal_2025,han2025adaptersalteringllmvocabularies,minixhofer2024zero} to obtain a distilled LLM whose vocabulary excludes gibberish tokens; this requires weight updates but shrinks the embedding matrix and the overall parameter count. Mapping deletion is preferable when query volume is low; vocabulary reduction pays an upfront distillation cost but lowers per-request serving cost at scale.

\paragraph{Implication on stakeholders of secret leakage.}
For \textit{online service providers} who provide secrets to users, they should be aware of the risk induced by tokenizers. For developers of code LLMs, they should proactively adopt the above mitigation strategy to mitigate the risk.
For \textit{academic researchers}, we advocate for red teaming strategies to understand every aspect of secret leakage. Such open questions include:  how can an adversary \textit{proactively} and \textit{effectively} exploit CLLMs to extract secrets? We leave these for future explorations.

\subsection{Implications on Tokenizer Design}

Our study surfaces two structural drawbacks of standard BPE tokenizers:

1.	\textbf{Limited flexibility.} BPE vocabulary is fixed prior to LLM pre-training; once fixed, Code LLM trainers should strictly follow the vocabulary to segment words into subwords, and adding or removing vocabulary items without full model re-training is challenging and under-explored. 

2.	\textbf{Sub-optimal compression utility under train-test distribution shift.} BPE has its historical origins in text compression, and it was brought to language models in 2016, since the community believes that compression boosts the performance of LLMs~\cite{huang2024compression}.  On LLMs, BPE is a heuristic-based algorithm to compress on the \textit{training} distribution. Therefore, its compression rate on specific \textit{test} distributions may degrade. In fact, we expect such a train-test distribution shift to exist for every downstream task of (Code) LLMs. Therefore, degradation of compression may affect downstream performance and the robustness of CLLMs. Our study presents a corner case for these distribution shifts, and shows the \textit{weird} behavior of BPE.

In summary,  BPE harms downstream task performance, and such defects may not be easily mitigated due to limited flexibility in vocabulary. To effectively adapt CLLMs to different downstream tasks, there have been a lot of efforts on CLLM post-training, agentic AI. We regard that an under-appreciated direction is promising: \textit{tokenizer adaptation}.

As a final remark, most evaluations of BPE to date are empirical, and the reasons for its good practical performance are not well understood across the AI community~\cite{kozma2024theoretical}. We then call for principled theoretical investigations towards BPE for academic researchers.

\section*{Limitations}
Limited by space, this paper fails to enumerate all choices of CLLMs since there are too many. We selected a popular subset of them. Besides, among all subword-based tokenization, we focus on BPE, since it is the dominant tokenization strategy in the community. We left further investigation on  Unigram and WordPiece for future research.

%% file: sections/ack.tex
\section*{Acknowledgment}
The work described in this paper was supported by the Research Grants Council of the Hong Kong Special Administrative Region, China (No. CUHK 14209124) of the General Research Fund, and RGC Grant for Theme-based Research Scheme Project (RGC Ref. No. T43-513/23-N).

%% file: sections/appendix.tex
\appendix

\section{Ethics considerations}
We contacted the authors in \citet{huang2025entropy} to obtain the secret dataset described in Section \ref{rq}. We use the secrets data solely for  statistical computing (e.g., entropy). We refrain from utilizing the sensitive information at any level.

\section{Example calculation of entropy}
\label{appdx}
Following the discussions in Section~\ref{bg_secret}, we consider the GitHub personal access token with regular expression \verb|ghp_[a-zA-Z0-9]{36}|. Assume the vocabulary is $\{A,\dots,Z,a,\dots,z,0\dots 9,\_\}$. Each char from the randomized part has an equal expected number of appearance $36\cdot \frac{1}{62}= \frac{18}{31}.$For $``g, h, p''$, the expected number of appearances is $1+\frac{18}{31}=\frac{49}{31}$. For ``$\_$'', the expected number of appearances is $1$. Therefore, for each of $g,h,p$:

  $$
    p(g)=p(h)=p(p)
    =\frac{49/31}{40}
    =\frac{49}{1240}
    \approx0.03952.
  $$
. For “\_”:

  $$
    p(\_) = \frac{1}{40}
           =0.02500.
  $$
. For each of the 59 “other” symbols:

  $$
    p(\text{other})
    =\frac{18/31}{40}
    =\frac{18}{1240}
    \approx0.01452.
  $$

Assume the base-2 entropy, the overall entropy is

\begin{align*}
H &= -\sum_i p_i \log p_i \\
  &= -\Bigl[\,3\cdot\frac{49}{1240}\log\frac{49}{1240}
     + \frac{31}{1240}\log\frac{31}{1240} \\
  &\qquad\qquad
     + 59\cdot\frac{18}{1240}\log\frac{18}{1240} \Bigr] \\
  &= 5.915 \quad \text{bits}.
\end{align*}

For the sample space size (i.e., vocabulary size) 63, the maximal entropy is achieved by a uniform distribution. 
$$
    H_{\max}=-\sum_i \frac{1}{63} \log \frac{1}{63} = 5.977. 
$$

The normalized entropy is
$$
H / H_{\max} = 5.915/5.977 = 0.9896 
$$

Through the example, we learn that a GitHub personal access token achieves around 99\% of the maximal entropy. 

\section{Character-wise Tokenization as a Mitigation}
\label{appdx:charwise}

\begin{figure*}[htbp]
    \centering
    \includegraphics[width=1\linewidth]{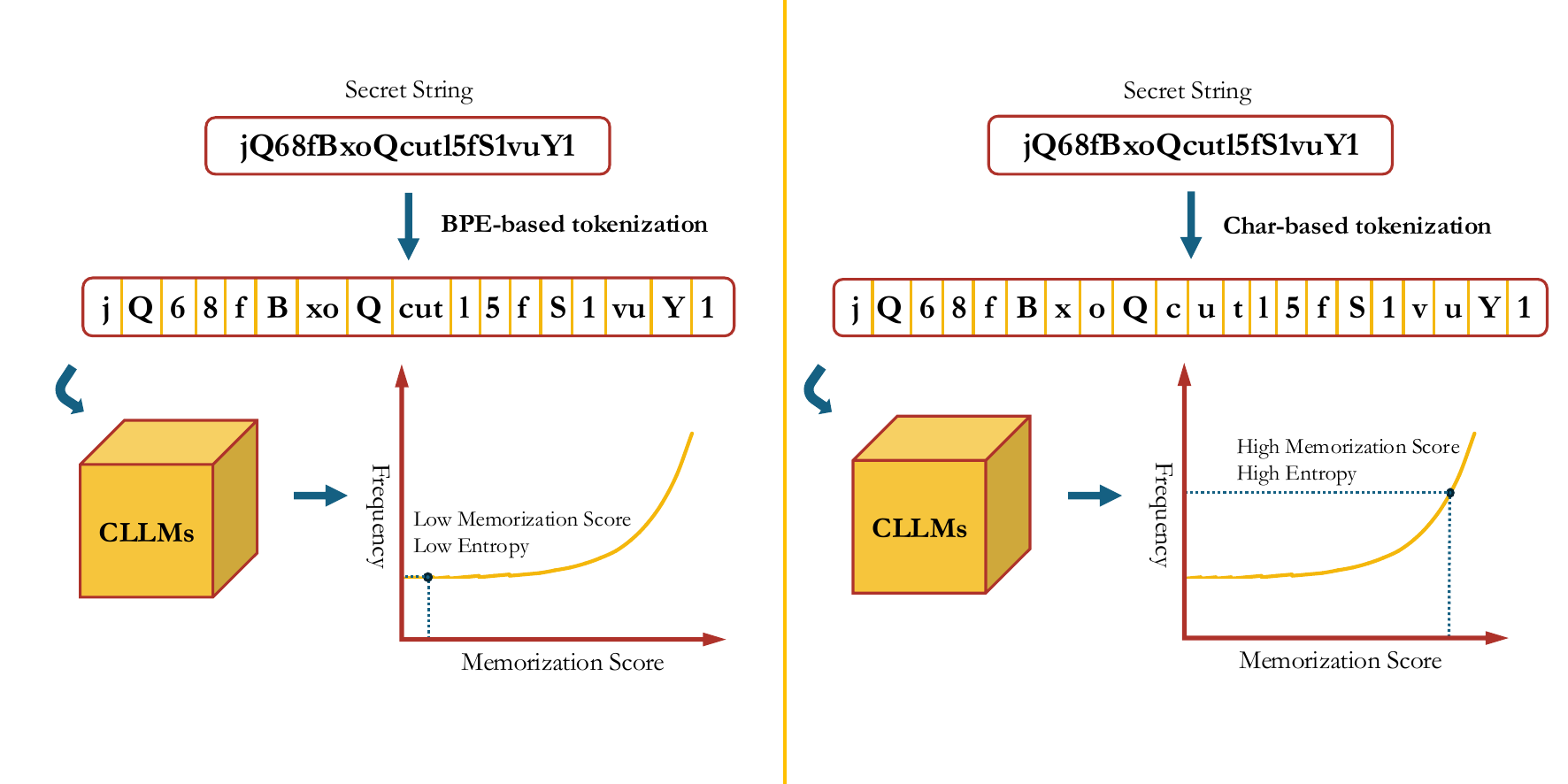}
    \caption{Mitigation Strategy Visualized}
    \label{fig:mitigation}
\end{figure*}

Gibberish bias is grounded on the discrepancy of how secret strings are processed at the design stage (char-level) and the CLLM inference stage (token-level). Therefore, a promising solution to address this gap is to force character-wise tokenization for code secrets. Recent research~\cite{vieira2025from} demonstrates that token-level models can be reinterpreted at the character level through a search-based alignment algorithm. By reconstructing conditional distributions over characters and employing beam search pruning for efficiency, this approach provides a principled means to approximate character-level behavior, thereby mitigating discrepancies in model processing. Figure~\ref{fig:mitigation} illustrates the overall idea.

The strategy aligns the discrepancy between the two stages. Therefore, we expect it to eliminate gibberish bias. The strategy also draws inspiration from digit-wise tokenization on integers in general-purpose LLMs, adopted in Llama-1~\cite{touvron2023llama}, Llama-2~\cite{llama2}, and Mistral~\cite{mistral7b}, aiming to explore arithmetic-related capabilities of LLMs.

\section{Tokenization Heatmaps on Unigram-based Tokenizers}
\label{appdx:unigram_heatmap}

\begin{figure}[htbp]
    \centering
    \begin{subfigure}[t]{0.49\linewidth}
        \includegraphics[width=\linewidth]{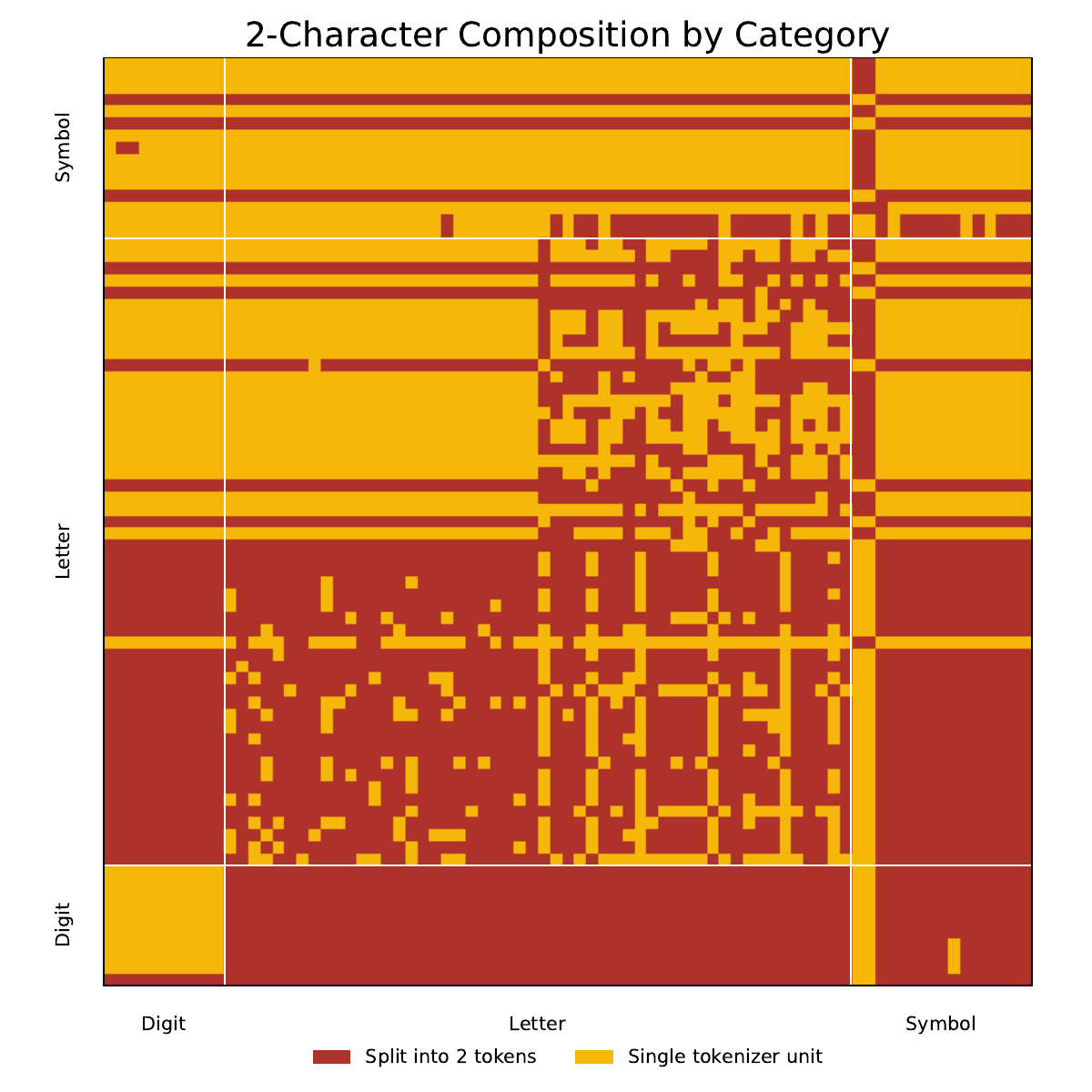}
    \end{subfigure}%
    \hfill
    \begin{subfigure}[t]{0.49\linewidth}
        \includegraphics[width=\linewidth]{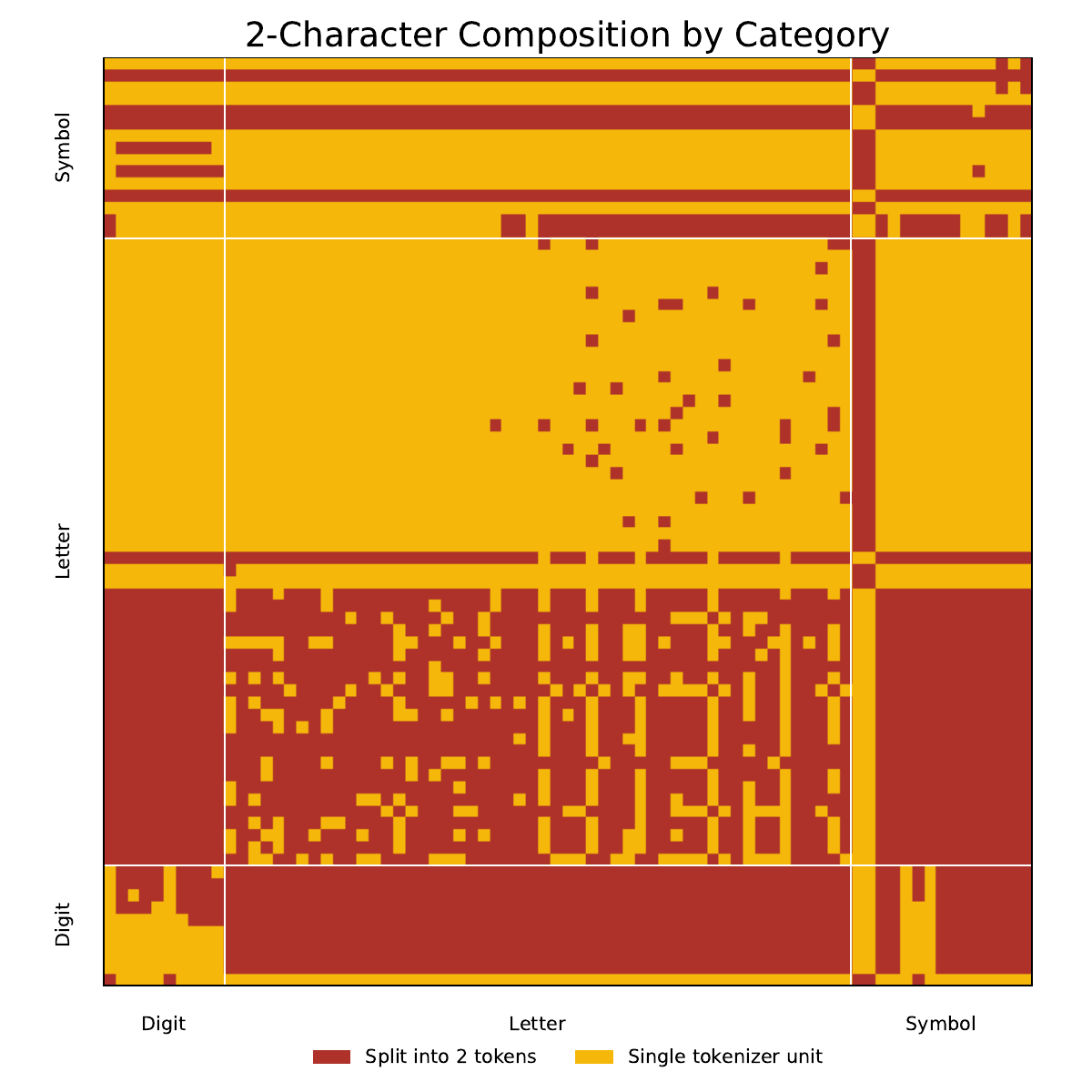}
    \end{subfigure}
    \caption{2-char sub-string tokenization on unigram-based tokenizers: XLNet (left) and T5 (right). Axes and color encoding follow Figure~\ref{fig:2-char}.}
    \label{fig:2-char-unigram}
\end{figure}

To check whether the counterintuitive tokenization behavior is specific to BPE, we repeat the 2-char enumeration experiment from Section~\ref{case_study} on two LLMs that adopt Unigram-based tokenization~\cite{kudo-richardson-2018-sentencepiece}: XLNet and T5. Figure~\ref{fig:2-char-unigram} shows that similar non-uniform token-splitting patterns persist, indicating that gibberish bias is not an artifact of BPE alone but a more general consequence of data-driven subword tokenization.